\documentclass[prd,aps,preprint,preprintnumbers,showpacs,nofootinbib,superscriptaddress]{revtex4}

\usepackage{amsmath}
\usepackage{amssymb}
\usepackage{graphicx}
\usepackage{txfonts}

\newcommand{\sla}[1]{{#1\hskip-0.45em/}}
\newcommand{\SU}{\textrm{SU}}
\newcommand{\SO}{\textrm{SO}}

\begin{document}


\preprint{RIKEN-TH-67}
\preprint{hep-th/0603146}

\title{%
Higher Order Terms of Improved Mean Field Approximation 
for \\ IIB Matrix Model 
and Emergence of Four-dimensional Space-time
}

\author{T.~Aoyama}
\affiliation{Theoretical Physics Laboratory, RIKEN, Wako, 351-0198, Japan }

\author{H.~Kawai}
\affiliation{Theoretical Physics Laboratory, RIKEN, Wako, 351-0198, Japan }
\affiliation{Department of Physics, Kyoto University, Kyoto, 606-8502, Japan }

\begin{abstract}
The spontaneous breakdown of $\SO(10)$ symmetry of the IIB matrix model 
has been studied by using the improved mean field approximation (IMFA). 
In this report, the eighth-order contribution to the improved 
perturbative series is obtained, which involves evaluation of 
20410 planar two-particle irreducible vacuum diagrams. 
We consider $\SO(d)$-preserving configurations as ansatz $(d=4,7)$. 
The development of \textit{plateau}, the solution of self-consistency 
condition, is seen in both ansatz. 
The large ratio of the space-time extent of $d$-dimensional part 
against the remaining $(10-d)$-dimensional part is obtained for 
$\SO(4)$ ansatz evaluated at the representative points of the plateau. 
It would be interpreted as the emergence of four-dimensional 
space-time in the IIB matrix model. 
\end{abstract}




\maketitle

\section{Introduction}

Superstring theory is supposed to provide a unified microscopic 
description of the universe including gravity. 
Recent progress revealed that many perturbative vacua thus far 
examined are related to each other, which enforces us to pursue 
non-perturbative formulation of superstring theory. 
One of those constructive definitions is the IIB matrix model 
\cite{ikkt} 
formulated in a form of the large-$N$ reduced model of the 
ten-dimensional supersymmetric $\SU(N)$ Yang-Mills theory. 
A characteristic feature of this model is that 
the eigenvalue distribution of ten bosonic matrices is interpreted 
as space-time, 
\textit{i.e.}, the space-time itself is treated as a dynamical variable 
of the model. 
It opens us a way toward explaining the origin of our four-dimensional 
space-time in the context of superstrings 
\cite{ikkt-prop}.

In this report, we consider the possibility of spontaneous breakdown 
of the original $\SO(10)$ symmetry 
through the dynamics of the IIB matrix model. 
It is examined by the anisotropy of eigenvalue distributions of 
the matrices. In this regard, the extents of space-time measured 
by the second moment of the matrices vary according to directions. 
It may occur that the $d$-dimensional subgroup of the rotational 
symmetry stays intact, which reflects that $d$-dimensional space-time 
emerges as the vacuum of the IIB matrix model. 

Exploring non-perturbative dynamics of a model 
is not an easy task in general. 
Monte Carlo simulation, a powerful tool that has been 
successfully applied to various physical models, 
is plagued by the sign problem in this case, 
because the action has a complex phase originated from 
the fermionic part.\footnote{%
A new technique called the factorization method is proposed 
to resolve the complex action problem in the Monte Carlo simulations 
\cite{factorization-method}.}
Whilst it is suggested that the complex phase plays crucial 
role in the spontaneous breakdown of Lorentz symmetry
\cite{phase}.

Here, we instead exploit a technique called the improved mean field 
approximation (IMFA). 
It is one of variational methods capable of exploring 
non-perturbative solutions of the model. 
It was developed 
in Ref.~\cite{imfa}, 
and first applied to IIB matrix model 
in Ref.~\cite{imfa-ikkt-3rd}, 
in which a systematic improvement to higher order terms was 
proposed and evaluated up to third order. 
The mechanism of spontaneous breakdown of rotational symmetry 
was further examined with the simplified models \cite{imfa-mm}. 
The analysis of the IIB matrix model by the IMFA method 
was proceeded to incorporate even higher-order contributions, 
up to fifth order 
in Ref.~\cite{imfa-ikkt-5th} 
and up to seventh order 
in Ref.~\cite{imfa-ikkt-7th}.
In the latter work the automated calculation procedure was also 
developed. 

The series of works showed that the spontaneous breakdown of 
Lorentz symmetry would occur to result in the emergence of 
the four-dimensional space-time. 
It was also observed that the extent of space-time of 
the four-dimensional part is larger than 
that of the remaining six-dimensional part, which is suggestive of 
the compactification scenario of string theory realized within the 
context of non-perturbative dynamics of the IIB matrix model. 

In the IMFA prescription, the Gaussian terms with auxiliary parameters 
are introduced into the action of the original model, which may be 
considered as mean fields in the mean field approximation. 
The self-consistency condition for those parameters is given by the 
principle of \textit{minimal sensitivity} 
\cite{pms} 
as a guide line. 
It should be realized as a region of parameter space denoted 
as \textit{plateau}. 
The development of plateau provides the indication whether or not 
the approximation is working well. 

In the present report, we proceed with the IMFA analysis 
of the IIB matrix model up to the eighth-order contribution. 
Since we are particularly interested in the spontaneous breakdown 
of $\SO(10)$ symmetry, we restrict ourselves to two particular 
choices of breaking patterns, namely, $\SO(4)$ ansatz and $\SO(7)$ 
ansatz, in which four- (seven-) dimensional rotational symmetry 
stays intact. 
By incorporating higher-order terms we expect a clearer signal 
for the emergence of plateau, as well as better estimates of 
the free energy and the ratio of the space-time extent 
in the true vacuum of the model. 

The paper is organized as follows. 
In Section~\ref{sec:2}, we provide a description of the model 
and the application of the IMFA analysis in a concrete procedure. 
The particular choices of configurations to be examined are 
given in Section~\ref{sec:3}. 
In Section~\ref{sec:4} we present the results. 
Section~\ref{sec:5} is devoted to the conclusion and discussions.

\section{IIB Matrix Model and IMFA Analysis\label{sec:2}}

\subsection{IIB Matrix Model\label{sec:2-1}}

The model we are considering here is the IIB matrix model defined 
by the partition function, 
\begin{equation}
	Z = \int\!dA\,d\psi\, e^{-S} \,,
	\quad
	S = \frac{1}{g_0^{\phantom{0}2}}\,{\rm Tr}\,
	\left\{
		-\frac{1}{4} \bigl[ A_\mu, A_\nu \bigr]^2 
		-\frac{1}{2} \bar{\psi} \Gamma^\mu \bigl[ A_\mu, \psi \bigr]
	\right\} \,,
\label{eq:action}
\end{equation}
where bosonic variables $A_\mu$ and fermionic variables $\psi_\alpha$ 
are both traceless $N \times N$ Hermite matrices. 
$A_\mu\ (\mu=1,\dots,10)$ transforms as a vector under $\SO(10)$ rotation, 
while $\psi_\alpha\ (\alpha=1,\dots,16)$ transforms as a left-handed spinor. 
The action has symmetry under the $\textrm{U}(N)$ matrix rotation, 
ten-dimensional Lorentz symmetry, and the type IIB supersymmetry. 

The coupling constant $g_0$ may be absorbed by the rescaling of 
the fields. Then the action takes the following form, 
\begin{equation}
	S = N\,{\rm Tr} 
	\left\{
		-\frac{1}{4} \bigl[ A_\mu, A_\nu \bigr]^2 
		-\frac{1}{2} \bar{\psi} \Gamma^\mu \bigl[ A_\mu, \psi \bigr]
	\right\} \,,
\label{eq:action2}
\end{equation}
where we have chosen $g_0^{\phantom{0}2} N=1$, and 
$A_\mu$ and $\psi_\alpha$ denote the rescaled fields.
We will examine the large-$N$ behaviour of the model. 

The extent of space-time along a certain direction $\mu$ is 
defined by the second moment of the matrix: 
\begin{equation}
	R_\mu^{\ 2} = \frac{1}{N}
	\left\langle {\rm Tr}\ A_\mu^{\ 2} \right\rangle
	\,.
\label{eq:sp-extent}
\end{equation}

\subsection{Improved Mean Field Approximation\label{sec:2-2}}

The action of the IIB matrix model (\ref{eq:action2}) does not 
have quadratic terms, and therefore 
the ordinary perturbation theory is not directly applicable. 
To such cases we employ a technique called 
the improved mean field approximation (IMFA) that 
is elucidated in the following. 

We introduce a quadratic term $S_0$ with arbitrary parameters 
collectively denoted by $m_0$ 
and deform the original action as 
\begin{equation}
	S\,\longrightarrow\,S_0 + \lambda (S - S_0) \,, 
\label{eq:action_deform}
\end{equation}
where $\lambda$ is a nominal parameter that should be taken to $1$. 
The second term of the deformed action (\ref{eq:action_deform}) may 
be considered as an interaction term with a coupling constant $\lambda$. 
So, we can formulate a perturbation theory and obtain 
a power series expansion with reference to $\lambda$. 
The series up to some finite order $n$ would provide an $n$th 
order approximation of the original model at $\lambda=1$. 
By this procedure, a formal perturbative expansion of a model 
can be constructed even when the model does not have quadratic terms, 
for example, as in the case of the IIB matrix model. 

The perturbative series thus obtained depends on the arbitrary 
parameters $m_0$. 
We have to determine the values of those parameters by some means. 
We adopt here the principle of minimal sensitivity 
\cite{pms} 
as a guiding principle; 
the \textit{true} value of a physical quantity should be given 
when it depends least on the arbitrary parameters. 
It is because the deformation (\ref{eq:action_deform}) 
of the action becomes trivial by construction 
if $\lambda$ is taken to 1, and the result is independent of 
those artificially introduced parameters. 
The dependence is brought in due to the truncation at the finite 
order of perturbation. 
Therefore, if a region of parameter space exists in which 
the series becomes almost constant, there the parameter dependence 
vanishes effectively, 
and the true value should be reproduced. 
We call this region ``plateau''. 
It has been tested and shown to work on a number of systems 
\cite{imfa,imfa-ikkt-3rd,imfa-ikkt-5th,imfa-ikkt-7th,imfa-ising,imfa-proc,imfa-mm}.

The concept of plateau above is rather obscure, 
and thus we need a more concrete criterion for the distinction of 
plateau in such a manner that the ambiguity due to subjectivity 
of recognition be excluded as much as possible. 
It is usually seen that the series as a function of the arbitrary 
parameters fluctuates on and near the plateau about the true 
value, and it is accompanied by a number of extrema of the function. 
This leads to a practical criterion for identifying plateau 
by the accumulation of extrema of the series with reference to 
the parameters.\footnote{%
An alternative approach for the identification of plateau from 
the profile of functions with the help of histograms 
is proposed in Ref.~\cite{imfa-proc}.
}
It has been adopted in the previous works 
\cite{imfa-ikkt-5th,imfa-ikkt-7th}. 
The values of the series are estimated at the extrema as the 
representatives of the plateau, which should give good approximations 
if the series is convergent. 
We adopt this criterion in the present analysis.

\subsection{Application to IIB Matrix Model\label{sec:2-3}}

Let us proceed to the IIB matrix model and evaluate the 
free energy of the model by the IMFA method. 
We introduce a quadratic term $S_0$ of most generic form 
that preserves $\textrm{U}(N)$ symmetry, 
\begin{equation}
	S_0 = N\,{\rm Tr}
	\left\{
		\frac{1}{2} M_{\mu\nu} A_\mu A_\nu
		+ \frac{1}{2} m_{\mu\nu\rho} 
		\bar{\psi}\Gamma^{\mu\nu\rho}\psi
	\right\} 
	\,.
\label{eq:gaussian}
\end{equation}
Here, $M_{\mu\nu}$ and $m_{\mu\nu\rho}$ are arbitrary parameters. 
$M_{\mu\nu}$ are symmetric with $\mu$ and $\nu$, while 
$m_{\mu\nu\rho}$ are totally anti-symmetric with $\mu$, $\nu$ and 
$\rho$. 

The prescription of the IMFA method is alternatively formulated 
by starting with the ``massive'' theory defined 
by the action, 
\begin{equation}
	S^\prime = S_0(M,m) + S(\lambda) \,,
\label{eq:deform_ite}
\end{equation}
where $S(\lambda)$ is given by inserting a formal parameter 
$\lambda$ to the original action (\ref{eq:action2}) as 
\begin{equation}
	S = N\,{\rm Tr} 
	\left\{
		-\frac{\lambda}{4} \bigl[ A_\mu, A_\nu \bigr]^2 
		-\frac{\sqrt{\lambda}}{2} 
		\bar{\psi} \Gamma^\mu \bigl[ A_\mu, \psi \bigr]
	\right\} \,. 
\label{eq:action3}
\end{equation}
A physical quantity $f$ is evaluated by the perturbative series 
expansion in terms of the coupling constant $\lambda$ 
up to some finite order $n$: 
\begin{equation}
	f(\lambda; M, m) = \sum_{k=0}^{n}\,\lambda^k\,f_k(M, m) \,.
\label{eq:f_taylor}
\end{equation}
Then, we perform the following substitution of parameters, 
\begin{align}
	M_{\mu\nu} & \longrightarrow (1-\lambda) M_{\mu\nu} \,,
\nonumber \\
	m_{\mu\nu\rho} & \longrightarrow (1-\lambda) m_{\mu\nu\rho} \,,
\label{eq:parameter_shift2}
\end{align}
and rearrange the series in powers of $\lambda$. 
Finally, We disregard $O(\lambda^{n+1})$ terms 
and set the formal expansion parameter $\lambda$ to 1. 
Following those steps, we obtain the \textit{improved} series 
$\widetilde{f}(\lambda; m)$. 
It will be denoted as 
\begin{equation}
	f \longrightarrow \widetilde{f}(M, m) 
	= 
	f\left( \lambda;\,(1-\lambda)M, (1-\lambda)m \right)
	\Bigr|_{\lambda^n, \lambda\rightarrow 1} \,.
\label{eq:f_improve}
\end{equation}
As is seen in Eq.~(\ref{eq:parameter_shift2}), the model is shifted 
to ``massless'' case at $\lambda=1$, and 
thus the original model would be reproduced.

For a technical reason, we exploit the fact that the ordinary 
free energy is related to the two-particle irreducible (2PI) 
free energy by the Legendre transformation, 
the latter of which is expressed by the sum of 2PI vacuum diagrams 
\cite{Fukuda:1995im}. 
It is because the number of diagrams incorporated reduces 
drastically. 

We evaluate the sum of 2PI vacuum diagrams in terms of 
the exact propagators. 
Since we are interested in the large-$N$ limit, the diagrams to be 
evaluated are restricted to planar ones. 
The 2PI free energy $G$ is given by: 
\begin{align}
	G(C,u) / N^2 = 
	& \ 
	3(1+\log 2) 
	+ \left\{ 
	- \frac{1}{2}{\rm tr}\log C + \frac{1}{2}{\rm tr}\log \sla{u} 
	\right\} 
\nonumber \\
	+ & 
	\lambda \left\{
	- \frac{1}{2}\bigl({\rm tr}(C^2)-({\rm tr}C)^2\bigr)
	- \frac{1}{2}C_{\mu\nu}{\rm tr}(\sla{u}\Gamma^\mu\sla{u}\Gamma^\nu)
	\right\} 
	+ \dots \,,
\label{eq:expr_G}
\end{align}
where $C_{\mu\nu}$ and $\sla{u} = u_{\mu\nu\rho}\Gamma^{\mu\nu\rho}/3!$ 
are the exact propagators of $A_\mu$ and $\psi$, respectively: 
\begin{align}
	\langle (A_{\mu})_{ij} (A_{\nu})_{kl} \rangle 
	&= 
	\frac{1}{N} C_{\mu\nu} \delta_{il} \delta_{jk} \,, 
\\
	\langle (\psi_{\alpha})_{ij} (\psi_{\beta})_{kl} \rangle 
	&=
	\frac{1}{N}\frac{i}{3!} u_{[\mu\nu\rho]} 
	(\Gamma^{\mu\nu\rho}{\cal C}^{-1})_{\alpha \beta} 
	\delta_{il}\delta_{jk} \,.
\label{eq:exact_propagators}
\end{align}
$\cal C$ is the charge conjugation matrix. 
The additive constants are adjusted according to the definition in 
Ref.~\cite{imfa-ikkt-3rd}. 
We then perform the Legendre transformation to obtain the free 
energy $F$ by 
\begin{equation}
	F(M, m) = 
	\left.
	\left(
	G(C, u) 
	+ \frac{1}{2} \sum_{\mu,\nu} M_{\mu\nu} C_{\mu\nu} 
	- \frac{1}{2} \sum_{\mu,\nu,\rho} m_{\mu\nu\rho} u_{\mu\nu\rho} 
	\right)
	\,\right|_{C=C(M,m), u=u(M,m)} \,,
\label{eq:legendre}
\end{equation}
where $C(M,m)$ and $u(M,m)$ are determined by the solutions of 
the following relations: 
\begin{equation}
	\frac{\partial G(C,u)}{\partial C_{\mu\nu}} 
	+ \frac{1}{2} M_{\mu\nu} = 0 \,, 
	\quad
	\frac{\partial G(C,u)}{\partial u_{\mu\nu\rho}} 
	- \frac{1}{2} m_{\mu\nu\rho} = 0 \,. 
\label{eq:legendre_solve}
\end{equation}

The improved free energy $\widetilde{F}$ is obtained by 
applying the procedure (\ref{eq:f_improve}) to $F$. 
Then, we search for the extrema of $\widetilde{F}(M,m)$ 
with reference to the parameters $M_{\mu\nu}$ and $m_{\mu\nu\rho}$, 
and identify the plateau by the accumulation of them. 
The values of free energy and other physical quantities are 
evaluated at the extrema.

\section{Ansatz\label{sec:3}}

In the case of the IIB matrix model, a huge number of parameters are 
introduced along with the quadratic terms: 
10 real numbers for $M_{\mu\nu}$ (assumed to be diagonalized), 
and 120 for $m_{\mu\nu\rho}$. 
It may demand enormous efforts to explore the whole parameter 
space for the solutions of plateau condition. 
Instead, we restrict ourselves to particular configurations 
in which $\SO(d)$ subgroup of $\SO(10)$ stays intact. 
The explicit forms of configurations are chosen according to the 
guideline described 
in Ref.~\cite{imfa-ikkt-5th}. 
In this report we concentrate on $d=4$ and $d=7$ cases in particular. 

Since we evaluate 2PI free energy in the first step, we specify 
the forms of exact propagators in each \textit{ansatz}.

\vskip 1ex

\noindent
{\bf $\SO(7)$ ansatz.} 
The fermionic propagators are represented by the rank three 
anti-symmetric tensor $u_{\mu\nu\rho}$. 
A non-zero element of $u$ accompanies three-dimensional block 
by considering the permutation of indices, and thus 
the $\SO(10)$ symmetry breaks down to $\SO(7)\times\SO(3)$. 

In this case, the propagators take the following forms 
with three parameters, $c_1$, $c_2$, and $u$: 
\begin{equation}
C_{\mu\nu} = \left(
	\begin{array}{cccc|ccc}
	c_1 & & & \\
	 & \ddots %
	\begin{picture}(0,0)\put(10,0){\rotatebox{52}{\raisebox{0pt}[0pt][0pt]{\makebox[0pt]{$\left.\rule{0pt}{50pt}\right\}$}}}}\put(10,7){7}\end{picture}%
	& & \\
	 & & \ddots & \\
	 & & & c_1 \\
	\hline 
	 & & & & c_2 \\
	 & & & & & \ddots %
	\begin{picture}(0,0)\put(-2,8){\rotatebox{52}{\raisebox{0pt}[0pt][0pt]{\makebox[0pt]{$\left.\rule{0pt}{30pt}\right\}$}}}}\put(0,12){3}\end{picture}%
	 \\
	 & & & & & & c_2
	\end{array}
\right) \,,
\quad
\sla{u} = 
	u\,\Gamma^{8,9,10} \,.
\end{equation}

\vskip 1ex

\noindent
{\bf $\SO(4)$ ansatz.}
We assume that $\SO(4)$ symmetry is preserved. 
The remaining 
six-dimensional subspace is decomposed into two three-dimensional blocks. 
This leads to the pattern of breaking as 
$\SO(4)\times\SO(3)\times\SO(3)\times Z_2$. 
The extra $Z_2$ factor derives from the permutation among 
two $\SO(3)$ factors with the reversion of the first direction 
(to preserve parity in total).

In this case, the propagators are represented by the form given below 
with three parameters, $c_1$, $c_2$, and $u$: 
\begin{equation}
C_{\mu\nu} = \left(
	\begin{array}{ccc|cccc}
	c_1 & & \\
	 & \ddots %
	\begin{picture}(0,0)\put(-2,8){\rotatebox{52}{\raisebox{0pt}[0pt][0pt]{\makebox[0pt]{$\left.\rule{0pt}{30pt}\right\}$}}}}\put(0,12){4}\end{picture}%
	& \\
	 & & c_1 \\
	\hline 
	 & & & c_2 \\
	 & & & & \ddots %
	\begin{picture}(0,0)\put(10,0){\rotatebox{52}{\raisebox{0pt}[0pt][0pt]{\makebox[0pt]{$\left.\rule{0pt}{50pt}\right\}$}}}}\put(10,7){6}\end{picture}%
	 \\
	 & & & & & \ddots \\
	 & & & & & & c_2
	\end{array}
\right) \,,
\quad
\sla{u} = 
	\frac{u}{\sqrt{2}}
	\left( \Gamma^{5,6,7} + \Gamma^{8,9,10} \right) \,.
\end{equation}

\section{Results\label{sec:4}}
\begin{figure}
\caption{Distributions of extrema of the improved free energy in the parameter space for $\SO(4)$ ansatz (left) and $\SO(7)$ ansatz (right).\label{fig:extrema}}
\begin{tabular}{cc}
\includegraphics[scale=0.85]{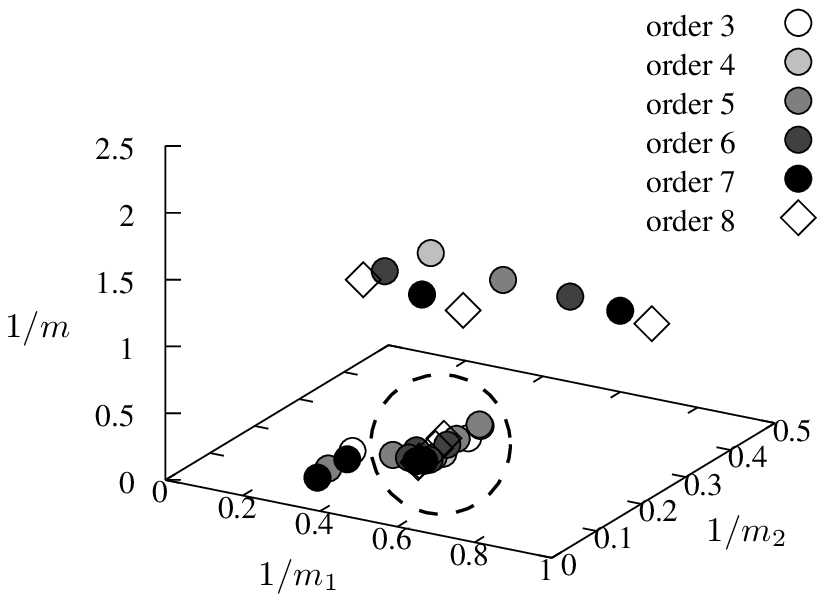} & 
\includegraphics[scale=0.85]{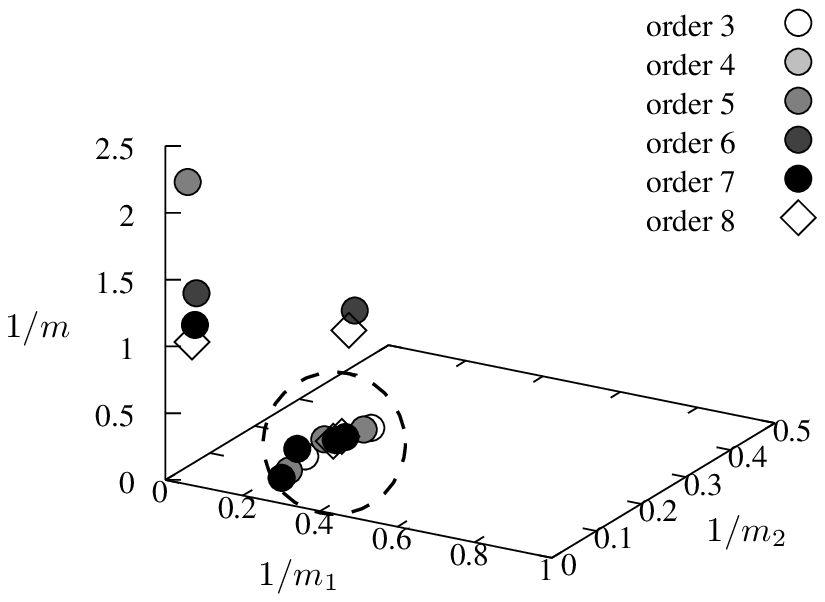} \\
\end{tabular}
\end{figure}
\begin{figure}
\caption{Free energy evaluated at the extrema for $\SO(4)$ ansatz (left) and $\SO(7)$ ansatz (right). Horizontal axis represents the order of approximation.\label{fig:f}}
\begin{tabular}{ccc}
\includegraphics[scale=.70]{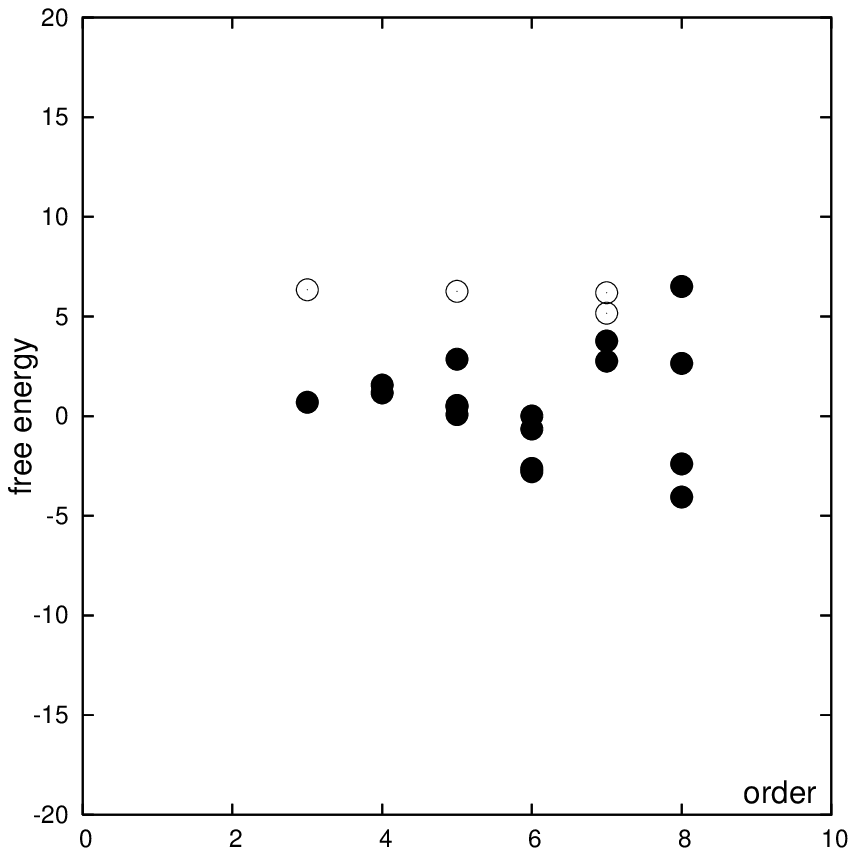} & 
\makebox[1em]{} & 
\includegraphics[scale=.70]{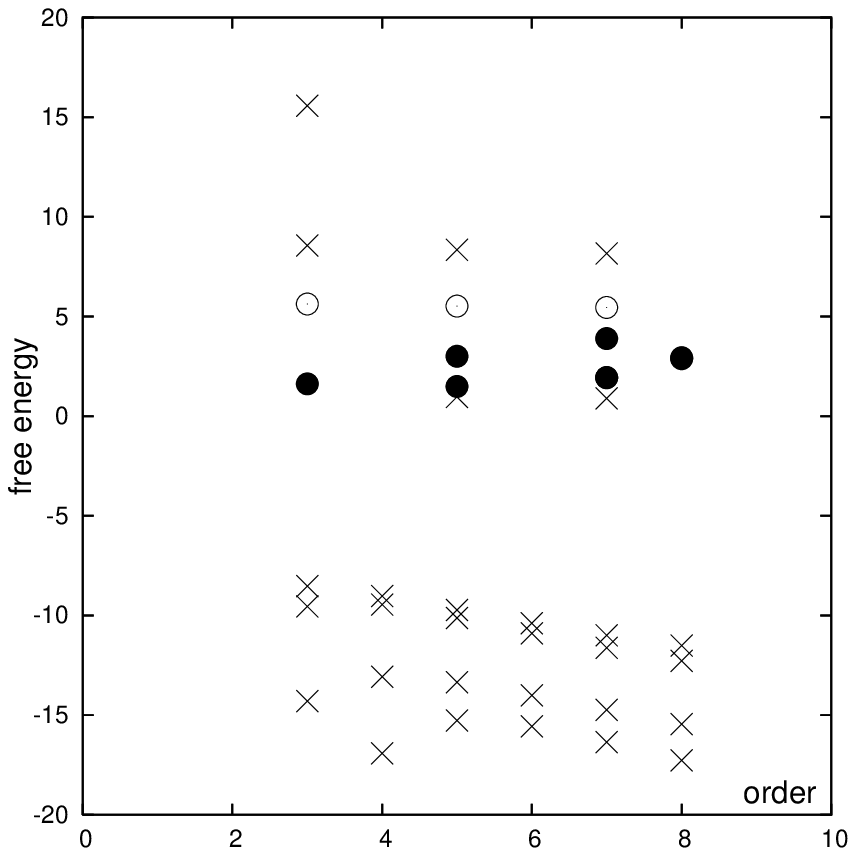} \\
{\small $\SO(4)$ ansatz} & & {\small $\SO(7)$ ansatz} \\
\end{tabular}
\end{figure}
\begin{figure}
\caption{Ratio of the extent of space-time evaluated at the extrema of the improved free energy for $\SO(4)$ ansatz (left) and $\SO(7)$ ansatz (right). Horizontal axis represents the order of approximation.\label{fig:r}}
\begin{tabular}{ccc}
\includegraphics[scale=.70]{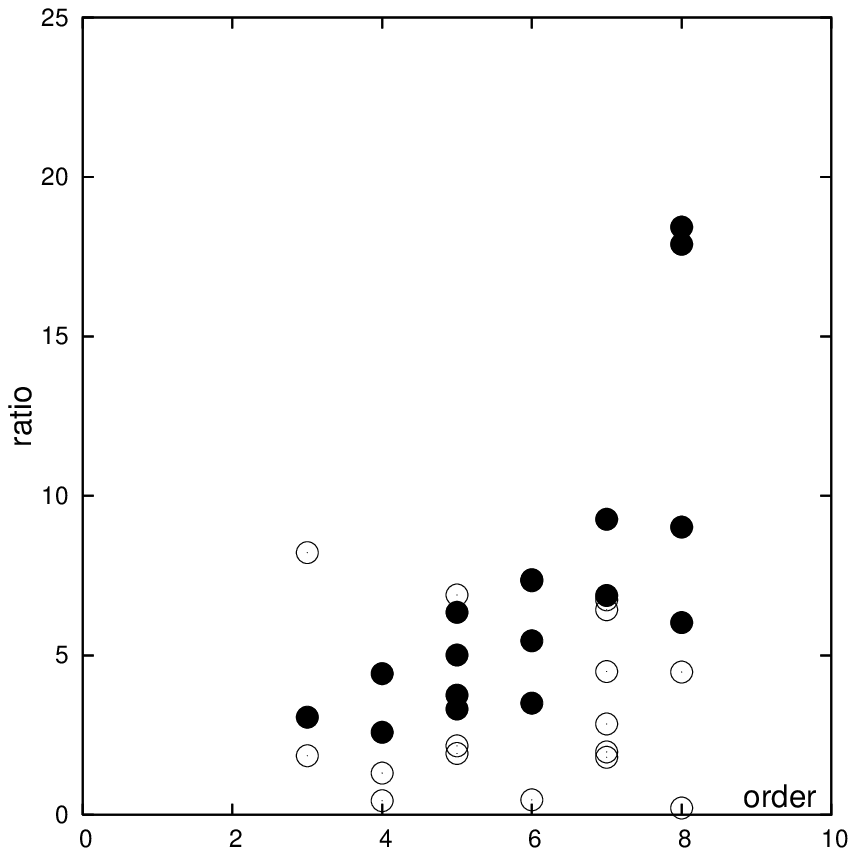} & 
\makebox[1em]{} & 
\includegraphics[scale=.70]{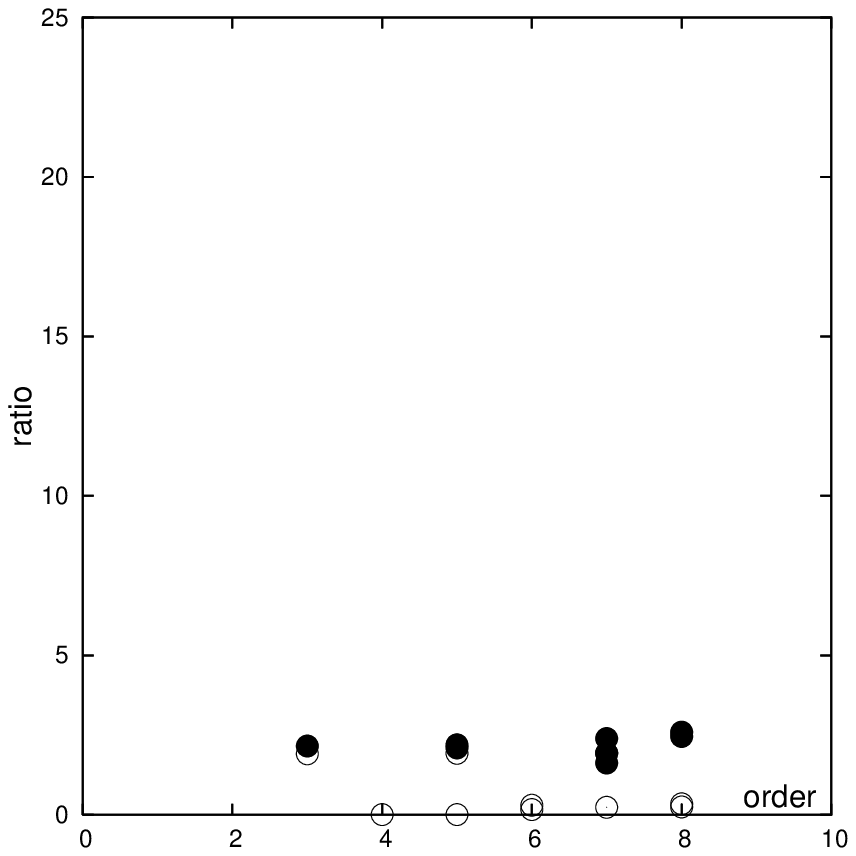} \\
{\small $\SO(4)$ ansatz} & & {\small $\SO(7)$ ansatz} \\
\end{tabular}
\end{figure}

We evaluated the free energy of the IIB matrix model for 
$\SO(4)$ ansatz and $\SO(7)$ ansatz by the IMFA method up to eighth order 
of series expansion. 
The number of planar 2PI diagrams are shown in 
Table~\ref{tbl:numdiagram}. 
In this report, the eighth-order contribution consisting of 
20410 distinct diagrams are newly evaluated. 
The automated procedure to generate and evaluate the set 
of diagrams has been developed in 
Ref.~\cite{imfa-ikkt-7th}, 
which is used in the present study with slight extensions. 
\begin{table}
\caption{Number of planar 2PI vacuum diagrams of (``massive'') 
IIB matrix model.\label{tbl:numdiagram}}
\begin{tabular}{crc}
\hline \hline 
\makebox[5em]{Order} & 
Number of diagrams & 
\makebox[1em]{} \\
\hline 
0th & 2 &\\
1st & 2 &\\
2nd & 2 &\\
3rd & 4 &\\
4th & 12 &\\
5th & 49 &\\
6th & 321 &\\
7th & 2346 &\\
8th & 20410 &\\
\hline \hline 
\end{tabular}
\end{table}

The improved free energy is obtained by following the procedure described 
in Sec.~\ref{sec:2} through the 2PI free energy $G$. 
Then, we search for the extrema of the improved free energy in the 
space of parameters $m_1$, $m_2$, and $m$ 
(dual to $c_1$, $c_2$, and $u$, respectively). 
The extrema of free energy for the eighth-order improved series are 
listed in Table~\ref{tbl:numerical}. 
We also confirmed that all the extrema of lower orders found in 
the former works \cite{imfa-ikkt-5th,imfa-ikkt-7th} 
are reproduced except one in seventh order of $\SO(4)$ ansatz. 

The distributions of extrema in the parameter space are plotted 
in Fig.~\ref{fig:extrema}. 
It is observed that a small region exists for both ansatz in which 
a number of extrema are found close to each other 
(shown in a dashed circle). 
It is recognized that extrema of different orders as well as those 
of respective orders belong to this accumulation. 
There are also extrema forming a line which would be considered 
as overshoots 
that characteristically appear around the edge of a plateau. 

In Fig.~\ref{fig:f}, the values of the free energy evaluated 
at the extrema are plotted 
for $\SO(4)$ and $\SO(7)$ ansatz against the order of the 
IMFA analysis. 
The bullets ($\bullet$) correspond to the extrema that belong to 
the accumulation, while the circles ($\circ$) 
correspond to the other extrema. 
The cross marks ($\times$) represent the unphysical extrema 
at which $c_1 < 0$ or $c_2 < 0$. 

Next, we consider what shapes of configurations are realized as 
the solutions. 
It is examined by evaluating the extent of space-time 
defined as the moments of the eigenvalue distributions: 
\begin{align}
	R^2 &= \frac{1}{N} \Bigl\langle {\rm Tr} A_{1} A_{1} \Bigr\rangle 
		= \frac{\partial F}{\partial m_1} 
		\biggr|_{\mbox{\scriptsize improved}} \,, 
\nonumber \\
	r^2 &= \frac{1}{N} \Bigl\langle {\rm Tr} A_{10} A_{10} \Bigr\rangle 
		= \frac{\partial F}{\partial m_2} 
		\biggr|_{\mbox{\scriptsize improved}} \,.
\end{align}
Here, the notation $\bigr|_{\mbox{\scriptsize improved}}$ denotes 
the application of the IMFA procedure to the series on the left. 

The values of $R$, $r$ and the ratio $\rho = R/r$ evaluated at the 
extrema of free energy are also listed in Table~\ref{tbl:numerical} 
for each ansatz. 
The ratio is plotted in Fig.~\ref{fig:r} for the order of approximation 
(the marks are the same as those of free energy). 
They show distinctive behaviour between $\SO(4)$ and $\SO(7)$ ansatz. 
In $\SO(7)$ case the ratio stays at $\rho \simeq 2.5$, while in $\SO(4)$ 
case it grows larger as the order increases. 
It should be remarked that we obtained the ratio 
of the space-time extent in $\SO(4)$ ansatz 
at an order of magnitude larger than 
the isotropic configuration, which has not been seen until 
the present eighth-order calculation. 

\begin{table}
\caption{Numerical values.\label{tbl:numerical}}
\begin{tabular}{c||r|r|r|r}
\hline \hline 
Ansatz  & 
\makebox[7em]{Free energy} & 
\makebox[7em]{$R^2$} & 
\makebox[7em]{$r^2$} & 
\makebox[7em]{$\rho$} \\
\hline 
$\SO(4)$ 
 & 2.646862 & 11.400026 & 0.033561 & 18.430393 \\
 & -4.056151 & 11.234590 & 0.035111 & 17.887798 \\
 & -2.393029 & 6.430896 & 0.078981 & 9.023488 \\
 & 6.507018 & 6.249471 & 0.171968 & 6.028345 \\
 & -29774.524354 & 165.895937 & 8.280539 & 4.475985 \\
 & -3610.819690 & 23.506193 & 532.358893 & 0.210130 \\
\hline 
$\SO(7)$ 
 & 2.885793 & 0.831046 & 0.123696 & 2.591994 \\
 & 2.931477 & 0.898519 & 0.148787 & 2.457429 \\
 & -52.619043 & 1.116045 & 9.571141 & 0.341475 \\
 & -108.828989 & 0.397967 & 6.643743 & 0.244747 \\
\hline \hline 
\end{tabular}

\end{table}

\section{Conclusion and Discussions\label{sec:5}}

We applied the improved mean field approximation (IMFA) to the IIB 
matrix model and evaluated up to eighth order contribution 
for configurations preserving $\SO(d)$ rotational symmetry as ansatz. 
We examined $d=4$ and $d=7$ cases in the present study. 

In order to solve the consistency condition for the arbitrary 
parameters called plateau condition, the stationary points 
(extrema) of the improved free energy are searched. 
It is observed that a region exists for each ansatz 
in which a number of extrema of eighth order 
as well as the extrema of different orders gather close to each other. 
It may be considered as a development of plateau. 
In the former works the discrepant behaviour was seen between even 
orders and odd orders of the improved series. 
There were no extrema found for $\SO(7)$ ansatz at even orders, 
and the free energy at extrema for $\SO(4)$ ansatz gave 
somewhat different sequence of values. 
Such discrepancy tends to be resolved with eighth-order terms taken 
into account. 
It would be consistent with the speculation that the 
result should be irrespective of the order of approximation, if 
the series is convergent and high enough orders are incorporated. 

We evaluated the extent of space-time of $d$-dimensional part 
and that of $(10-d)$-dimensional part in $\SO(d)$ ansatz. 
They are given by the moments of the eigenvalue distributions 
at the extrema of free energy as the representative 
points of plateau. 
The deviation of the ratio of the extents from 1 represents 
the anisotropy of space-time realized as a non-perturbative vacuum 
of the IIB matrix model, 
and thus provides an indication of the spontaneous breakdown of 
Lorentz symmetry. 
In the present work, the ratio of an order of magnitude 
larger than the isotropic configuration was obtained 
for $\SO(4)$ ansatz. 
It might suggest that the actual ratio realized in the true vacuum 
of the theory may possibly be infinite, 
which implies the emergence of extended four-dimensional universe 
with the remaining six-dimensional part being compactified, as 
seen in the universe. 

Still we do not have clear signal enough to infer 
the formation of plateau, and therefore 
no reasonable estimate for free energy nor other physical values 
would be assured at this stage. 
The configuration with smaller values of free energy should give 
dominant contribution in the whole configuration space, and 
it is supposed to be realized as our universe. 
In this regard, we can not yet definitely tell which of the ansatz 
would be plausible. 
More sophisticated scheme for identifying plateau and 
extracting the physical quantities would be required 
along with the effort to proceed to even higher order contributions.

\begin{acknowledgments}
The authors would like to thank 
M.~Hayakawa, 
S.~Kawamoto, 
T.~Kuroki, 
T.~Matsuo, 
J.~Nishimura, 
and 
Y.~Shibusa 
for valuable discussions and helpful comments. 
A part of calculation was performed by using the computational 
resources of the RIKEN Super Combined Cluster (RSCC). 
\end{acknowledgments}




\end{document}